
\voffset 0.1in
\hoffset 0.1in

\def\gboxit#1{\hbox{\vrule\vbox{\hrule\kern3pt\vtop
{\hbox{\kern3pt#1\kern3pt}
\kern3pt\hrule}}\vrule}}

\def\ttilde#1{\raise2ex\hbox{${\scriptscriptstyle(}\!
\sim\scriptscriptstyle{)}$}\mkern-16.5mu #1}
\def\dddots#1{\raise1ex\hbox{$^{\ldots}$}\mkern-16.5mu #1}
\def\siton#1#2{\raise1.5ex\hbox{$\scriptscriptstyle{#2}$}\mkern-16.5mu
#1}
\def\upleftarrow#1{\raise1.5ex\hbox{$\leftarrow$}\mkern-16.5mu #1}
\def\uprightarrow#1{\raise1.5ex\hbox{$\rightarrow$}\mkern-16.5mu #1}
\def\upleftrightarrow#1{\raise1.5ex\hbox{$\leftrightarrow$}\mkern-16.5mu
#1}
\def\bx#1#2{\vcenter{\hrule \hbox{\vrule height #2in \kern
#1in\vrule}\hrule}}

\def\squiggle#1{\lower1.5ex\hbox{$\sim$}\mkern-14mu #1}

\def\narrower{\advance\leftskip by\parindent \advance\rightskip
by\parindent}

\def\mbox#1#2{\vcenter{\hrule width#1in\hbox{\vrule height#2in
   \hskip#1in\vrule height#2in}\hrule width#1in}}
\def\eqsquare #1:#2:{\vcenter{\hrule width#1\hbox{\vrule height#2
   \hskip#1\vrule height#2}\hrule width#1}}
\def\inbox#1#2#3{\vcenter to #2in{\vfil\hbox to
#1in{$$\hfil#3\hfil$$}\vfil}}
\def\strutdepth{\dp\strutbox}
\def\marbul{\strut\vadjust{\kern-\strutdepth\specialbul}}
\def\specialbul{\vtop to \strutdepth{
    \baselineskip\strutdepth\vss\llap{$\bullet$\qquad}\null}}
\def\Bcomma{\lower6pt\hbox{$,$}}    
\def\bcomma{\lower3pt\hbox{$,$}}    

\def\sl{\scrsf}

\def\updots{\mathinner{\mskip 1mu\raise 1pt\hbox{.}
    \mskip 2mu\raise 4pt\hbox{.}\mskip 2mu
    \raise 7pt\vbox{\kern 7pt\hbox{.}}\mskip 1mu}}

\def\square{\kern1pt\vbox{\hrule height 1.2pt\hbox{\vrule width
1.2pt\hskip 3pt
   \vbox{\vskip 6pt}\hskip 3pt\vrule width 0.6pt}\hrule height
0.6pt}\kern1pt}
\def\ssquare{\kern1pt\vbox{\hrule height .6pt\hbox{\vrule width
.6pt\hskip 3pt
   \vbox{\vskip 6pt}\hskip 3pt\vrule width 0.6pt}\hrule height
0.6pt}\kern1pt}
\def\lege{\hbox{$ {     \lower.40ex\hbox{$>$}
                   \atop \raise.20ex\hbox{$<$}
                   }     $}  }

\def\rege{\hbox{$ {     \lower.40ex\hbox{$<$}
                   \atop \raise.20ex\hbox{$>$}
                   }     $}  }

\def\lapp{\hbox{$ {     \lower.40ex\hbox{$<$}
                   \atop \raise.20ex\hbox{$\sim$}
                   }     $}  }
\def\rapp{\hbox{$ {     \lower.40ex\hbox{$>$}
                   \atop \raise.20ex\hbox{$\sim$}
                   }     $}  }

\def\tridots{\hbox{$ {     \lower.40ex\hbox{$.$}
                   \atop \raise.20ex\hbox{$.\,.$}
                   }     $}  }
\def\Times{\times\hskip-2.3pt{\raise.25ex\hbox{{$\scriptscriptstyle|$}}}}

\def\rightonleft{\hbox{$ {     \lower.40ex\hbox{$\longrightarrow$}
                   \atop \raise.20ex\hbox{$\longleftarrow$}
                   }     $}  }

\def\pmb#1{\setbox0=\hbox{#1}%
\kern-.025em\copy0\kern-\wd0
\kern.05em\copy0\kern-\wd0
\kern-.025em\raise.0433em\box0 }
%
%
\font\fivebf=cmbx5
\font\sixbf=cmbx6
\font\sevenbf=cmbx7
\font\eightbf=cmbx8
\font\ninebf=cmbx9
\font\tenbf=cmbx10

\font\bfmone=cmbx10 scaled\magstep1

\font\sevenit=cmti7
\font\eightit=cmti8
\font\nineit=cmti9
\font\tenit=cmti10

\font\itmone=cmti10 scaled\magstep1

\font\fiverm=cmr5
\font\sixrm=cmr6
\font\sevenrm=cmr7
\font\eightrm=cmr8
\font\ninerm=cmr9
\font\tenrm=cmr10

\font\rmmone=cmr10 scaled\magstep1

\def\fontone{\def\rm{\fcm0\rmmone}%
  \textfont0=\rmmone \scriptfont0=\tenrm \scriptscriptfont0=\sevenrm
  \textfont1=\itmone \scriptfont1=\tenit \scriptscriptfont1=\sevenit
  \def\it{\fcm\itfcm\itmone}%
  \textfont\itfcm=\itmone
  \def\bf{\fcm\bffcm\bfmone}%
  \textfont\bffcm=\bfmone \scriptfont\bffcm=\tenbf
   \scriptscriptfont\bffcm=\sevenbf
  \tt \ttglue=.5em plus.25em minus.15em
  \normalbaselineskip=25pt
  \let\sc=\tenrm
  \let\big=\tenbig
  \setbox\strutbox=\hbox{\vrule height10.2pt depth4.2pt width\z@}%
  \normalbaselines\rm}



\font\ninerm=cmr9
\font\eightrm=cmr8
\font\sixrm=cmr6

\font\ninei=cmmi9
\font\eighti=cmmi8
\font\sixi=cmmi6
\skewchar\ninei='177 \skewchar\eighti='177 \skewchar\sixi='177

\font\ninesy=cmsy9
\font\eightsy=cmsy8
\font\sixsy=cmsy6
\skewchar\ninesy='60 \skewchar\eightsy='60 \skewchar\sixsy='60

\font\ninebf=cmbx9
\font\eightbf=cmbx8
\font\sixbf=cmbx6

\font\ninett=cmtt9
\font\eighttt=cmtt8

\hyphenchar\tentt=-1 
\hyphenchar\ninett=-1
\hyphenchar\eighttt=-1

\font\ninesl=cmsl9
\font\eightsl=cmsl8

\font\nineit=cmti9
\font\eightit=cmti8


\newskip\ttglue
\def\tenpoint{\def\rm{\fcm0\tenrm}%
  \textfont0=\tenrm \scriptfont0=\sevenrm \scriptscriptfont0=\fiverm
  \textfont1=\teni \scriptfont1=\seveni \scriptscriptfont1=\fivei
  \textfont2=\tensy \scriptfont2=\sevensy \scriptscriptfont2=\fivesy
  \textfont3=\tenex \scriptfont3=\tenex \scriptscriptfont3=\tenex
  \def\it{\fcm\itfcm\tenit}%
  \textfont\itfcm=\tenit
  \def\sl{\fcm\slfcm\tensl}%
  \textfont\slfcm=\tensl
  \def\bf{\fcm\bffcm\tenbf}%
  \textfont\bffcm=\tenbf \scriptfont\bffcm=\sevenbf
   \scriptscriptfont\bffcm=\fivebf
  \def\tt{\fcm\ttfcm\tentt}%
  \textfont\ttfcm=\tentt
  \tt \ttglue=.5em plus.25em minus.15em
  \normalbaselineskip=16pt
  \let\sc=\eightrm
  \let\big=\tenbig
  \setbox\strutbox=\hbox{\vrule height8.5pt depth3.5pt width\z@}%
  \normalbaselines\rm}

\def\ninepoint{\def\rm{\fcm0\ninerm}%
  \textfont0=\ninerm \scriptfont0=\sixrm \scriptscriptfont0=\fiverm
  \textfont1=\ninei \scriptfont1=\sixi \scriptscriptfont1=\fivei
  \textfont2=\ninesy \scriptfont2=\sixsy \scriptscriptfont2=\fivesy
  \textfont3=\tenex \scriptfont3=\tenex \scriptscriptfont3=\tenex
  \def\it{\fcm\itfcm\nineit}%
  \textfont\itfcm=\nineit
  \def\sl{\fcm\slfcm\ninesl}%
  \textfont\slfcm=\ninesl
  \def\bf{\fcm\bffcm\ninebf}%
  \textfont\bffcm=\ninebf \scriptfont\bffcm=\sixbf
   \scriptscriptfont\bffcm=\fivebf
  \def\tt{\fcm\ttfcm\ninett}%
  \textfont\ttfcm=\ninett
  \tt \ttglue=.5em plus.25em minus.15em
  \normalbaselineskip=11pt
  \let\sc=\sevenrm
  \let\big=\ninebig
  \setbox\strutbox=\hbox{\vrule height8pt depth3pt width\z@}%
  \normalbaselines\rm}

\def\eightpoint{\def\rm{\fcm0\eightrm}%
  \textfont0=\eightrm \scriptfont0=\sixrm \scriptscriptfont0=\fiverm
  \textfont1=\eighti \scriptfont1=\sixi \scriptscriptfont1=\fivei
  \textfont2=\eightsy \scriptfont2=\sixsy \scriptscriptfont2=\fivesy
  \textfont3=\tenex \scriptfont3=\tenex \scriptscriptfont3=\tenex
  \def\it{\fcm\itfcm\eightit}%
  \textfont\itfcm=\eightit
  \def\sl{\fcm\slfcm\eightsl}%
  \textfont\slfcm=\eightsl
  \def\bf{\fcm\bffcm\eightbf}%
  \textfont\bffcm=\eightbf \scriptfont\bffcm=\sixbf
   \scriptscriptfont\bffcm=\fivebf
  \def\tt{\fcm\ttfcm\eighttt}%
  \textfont\ttfcm=\eighttt
  \tt \ttglue=.5em plus.25em minus.15em
  \normalbaselineskip=9pt
  \let\sc=\sixrm
  \let\big=\eightbig
  \setbox\strutbox=\hbox{\vrule height7pt depth2pt width\z@}%
  \normalbaselines\rm}



\magnification=1200
\newbox\leftpage
\newdimen\fullhsize
\newdimen\hstitle
\newdimen\hsbody
\tolerance=600\hfuzz=2pt
\hoffset=0.1truein \voffset=0.1truein
\hsbody=\hsize \hstitle=\hsize
\font\titlefont=cmr10 scaled\magstep3
\font\secfont=cmbx10 scaled\magstep1

\def\nolabels{\def\eqnlabel##1{}\def\eqlabel##1{}\def\reflabel##1{}}
\def\writelabels{\def\eqnlabel##1{%
{\escapechar=` \hfill\rlap{\hskip.09in\string##1}}}%
\def\eqlabel##1{{\escapechar=` \rlap{\hskip.09in\string##1}}}%
\def\reflabel##1{\noexpand\llap{\string\string\string##1\hskip.31in}}}
\nolabels
\def\title#1{ \nopagenumbers\hsize=\hsbody%
\centerline{ {\titlefont #1} }%
\pageno=0}
\def\author#1{\vskip 1 truecm%
\centerline{{\sl #1}}%
\centerline{Center for Theoretical Physics}%
\centerline{Laboratory for Nuclear Science}%
\centerline{and Department of Physics}%
\centerline{Massachusetts Institute of Technology}%
\centerline{Cambridge, Massachusetts 02139}}
\def\abstract#1{\centerline{\bf ABSTRACT}\nobreak\medskip\nobreak\par #1}

\global\newcount\secno \global\secno=0
\global\newcount\meqno \global\meqno=1
\def\newsec#1{\global\advance\secno by1
\xdef\secsym{\ifcase\secno
\or I\or II\or III\or IV\or V\or VI\or VII\or VIII\or IX\or X\fi }
\global\meqno=1
\bigbreak\bigskip
\noindent{\secfont\secsym. #1}\par\nobreak\medskip\nobreak}
\xdef\secsym{}


\def\eqnn#1{\xdef #1{(\the\secno.\the\meqno)}%
\global\advance\meqno by1\eqnlabel#1}
\def\eqna#1{\xdef #1##1{\hbox{$(\the\secno.\the\meqno##1)$}}%
\global\advance\meqno by1\eqnlabel{#1$\{\}$}}
\def\eqn#1#2{\xdef #1{(\the\secno.\the\meqno)}\global\advance\meqno by1%
$$#2\eqno#1\eqlabel#1$$}
%
%
%

\def\eqnla#1#2#3{\xdef #2{(\the\secno.\the\meqno a)}
\xdef #1{(\the\secno.\the\meqno)}
$$#3\eqno#2\eqlabel#1\eqlabel#2$$}
\def\eqnlb#1#2{\xdef #1{(\the\secno.\the\meqno b)}
$$#2\eqno#1\eqlabel#1$$}
\def\eqnlc#1#2{\xdef #1{(\the\secno.\the\meqno c)}
$$#2\eqno#1\eqlabel#1$$}
\def\eqnld#1#2{\xdef #1{(\the\secno.\the\meqno d)}
$$#2\eqno#1\eqlabel#1$$}
\def\eqnle#1#2{\xdef #1{(\the\secno.\the\meqno e)}
$$#2\eqno#1\eqlabel#1$$}
\def\eqnlf#1#2{\xdef #1{(\the\secno.\the\meqno f)}
$$#2\eqno#1\eqlabel#1$$}
\def\eqnlbend#1#2{\xdef #1{(\the\secno.\the\meqno b)}\global\advance\meqno by1%
$$#2\eqno#1\eqlabel#1$$}
\def\eqnlcend#1#2{\xdef #1{(\the\secno.\the\meqno c)}\global\advance\meqno by1%
$$#2\eqno#1\eqlabel#1$$}
\def\eqnldend#1#2{\xdef #1{(\the\secno.\the\meqno d)}\global\advance\meqno by1%
$$#2\eqno#1\eqlabel#1$$}
\def\eqnleend#1#2{\xdef #1{(\the\secno.\the\meqno e)}\global\advance\meqno by1%
$$#2\eqno#1\eqlabel#1$$}
\def\eqnlfend#1#2{\xdef #1{(\the\secno.\the\meqno f)}\global\advance\meqno by1%
$$#2\eqno#1\eqlabel#1$$}
\def\eqnlgend#1#2{\xdef #1{(\the\secno.\the\meqno g)}\global\advance\meqno by1%
$$#2\eqno#1\eqlabel#1$$}
\global\newcount\ftno \global\ftno=1
\def\foot#1{{\baselineskip=12pt plus 1pt\footnote{$^{\the\ftno}$}{#1}}%
\global\advance\ftno by1}
\global\newcount\refno \global\refno=1
\newwrite\rfile
\def\ref#1#2{\the\refno\nref#1{#2}}
\def\nref#1#2{\xdef#1{\the\refno}%
\ifnum\refno=1\immediate\openout\rfile=refs.tmp\fi%
\immediate\write\rfile{\noexpand\item{#1\ }\reflabel{#1}#2}%
\global\advance\refno by1}
\def\addref#1{\immediate\write\rfile{\noexpand\item{}#1}}

\def\figures{\centerline{{\bf Figure Captions}}\medskip\parindent=40pt}
\def\fig#1#2{\medskip\item{Fig.~#1:  }#2}

\def\frac#1#2{{#1\over#2}}
\pageno=0
$^{\ref\kili{D. Kirzhnits, JETP Lett. {\bf 15}, 529 (1972);
D. Kirzhnits and A. Linde, Phys. Lett. {\bf B 42}, 471
(1972).}}$
$^{\ref\doja{L. Dolan and R. Jackiw, Phys. Rev. {\bf D 9}, 3320 (1974).}}$
$^{\ref\wei{S. Weinberg, Phys. Rev. {\bf D 9}, 3357 (1974).}}$
$^{\ref\kilib{D. Kirzhnits and A. Linde,
JETP Lett. {\bf 40}, 628 (1974).}}$

$^{\ref\mat{For recent reviews, see: A.D. Dolgov, Kyoto preprint
YITP/K-940 (1991); M.E. Shaposhnikov, CERN preprint CERN-TH.6304/91
(1991);
M. Dine, Santa Cruz preprint SCIPP 91/27 (1991).}}$

$^{\ref\smref{M.E. Shaposhnikov, Phys. Lett. {\bf B 277}, 324 (1992).}}$

$^{\ref\car{M.E. Carrington, Phys. Rev. {\bf D 45}, 2933 (1992).}}$

$^{\ref\firef{M. Dine, R.G. Leigh, P. Huet, A. Linde, and D. Linde,
Phys. Rev. {\bf D 46}, 550 (1992) and Phys. Lett. {\bf B 283}, 319
(1992).}}$

$^{\ref\arn{P. Arnold, Phys. Rev. {\bf D 46}, 2628 (1992).}}$

$^{\ref\hsu{S.D.H. Hsu, to
appear in the proceedings of the Yale/Texas Workshop on (B+L) Violation
(World Scientific).}}$

$^{\ref\jai{V. Jain,  MPI-Ph/92-41 (1992).}}$

$^{\ref\zwi{J.R. Espinosa, M. Quiros, and F. Zwirner,
CERN-TH-6451/92 (1992).}}$

$^{\ref\boy{C.G. Boyd, D.E. Brahm and S.D.H. Hsu,
CALT-68-1795/HUTP-92-A027/EFI-92-22 (1992).}}$

$^{\ref\corn{J.M. Cornwall, R. Jackiw, and E. Tomboulis,
Phys. Rev. {\bf D 10}, 2428 (1974).}}$

$^{\ref\dine{Dine et al. in Ref.[\firef] also pointed out  that the
one-loop improved term by itself leads to wrong counting.}}$

$^{\ref\pisa{The $T=0$ limit of Equation (3.16)
has been obtained previously using the Gaussian approximation in the
Schroedinger picture, which is equivalent to the present approximation.
See
S.-Y. Pi and M. Samiullah,
Phys. Rev. {\bf D 36}, 3128 (1987).}}$

$^{\ref\momo{See, for example, S. Coleman, R. Jackiw, and H.D. Politzer,
Phys. Rev. {\bf D 10}, 2491 (1974);
W.A. Bardeen and Moshe Moshe,
Phys. Rev. {\bf D 28}, 1372 (1983).}}$

\vfill
\eject
\vfill

\centerline{{\bf SELF-CONSISTENT IMPROVEMENT OF THE }}
\centerline{{\bf FINITE TEMPERATURE EFFECTIVE
POTENTIAL}\footnote{*}
{This work is
supported in part by funds provided by the U.S. Department of Energy (D.O.E.)
under contract $\#$DE-FG02-91ER40676. }}
\vskip 48pt
\bigskip

\centerline{G. Amelino-Camelia\footnote{**}
{internet address: gac@budoe.bu.edu}
and So-Young Pi}
\vskip 18pt
\baselineskip 12pt plus 0.2pt minus 0.2pt
\centerline{Department of Physics}
\centerline{Boston University}
\centerline{590 Commonwealth Ave.}
\centerline{Boston, Massachusetts 02215}
\vskip 4.0cm
\centerline{\bf ABSTRACT }
We present a self-consistent calculation of the finite temperature
effective potential for $\lambda \phi^4$ theory, using the composite
operator effective potential in which an infinite series of the leading
diagrams is summed up. Our calculation establishes the proper form of
the leading correction to the perturbative one-loop effective potential.

\vskip 1cm
\vfill
\noindent{BUHEP-92-26 \hfill November, 1992}
\eject
\vfill
\baselineskip 24pt plus 0.2pt minus 0.2pt
\newsec{Introduction}
Temperature induced symmetry-changing phase-transitions in quantum field
theory are important ingredients in cosmological scenarios. The
existence of the high-temperature phase transitions was
suggested by Kirzhnits and
Linde${[\kili]}$, and was shown quantitatively by
 Dolan and Jackiw${[\doja]}$,  Weinberg${[\wei]}$, and Kirzhnits and
Linde${[\kilib]}$, by calculating the finite temperature one-loop effective
potential. However, cosmological
scenarios often rely on the detailed nature of the phase transition, namely
whether it is of first or second order.
A more precise determination of the critical temperature and
the nature of the phase transition requires an analysis of higher-loop
contributions even when the coupling constants in the theory are very
small. Weinberg has argued, by using power counting, that the leading
contributions at very high temperatures come from all those loops with
superficial degree of divergence D$>1$. This implies that in order to
obtain more accurate information one must study infinite series of
certain classes of multi-loop diagrams in perturbation theory. For
example, in $\lambda \Phi^4$ scalar theory the leading contributions
come from the multi-loop graphs shown in Fig.1, which are
called {\it daisy} and {\it super-daisy}.
For this reason, Dolan and Jackiw, in their early paper, studied the
effect of these graphs on the temperature dependent effective mass.

Recently there has been much interest in the nature of the electroweak
phase transition due to the idea that the baryon asymmetry may be
generated at the electroweak scale if the transition is of first
order${[\mat]}$. Several authors have calculated the high
temperature effective potential in
the standard model and in the $\lambda \Phi^4$ theory, taking into
account
leading (and subleading) contributions from multi-loop diagrams, in
order to obtain a correct form of the high temperature effective
potential${[\smref,\boy]}$. Some of the authors have
calculated an ``improved''
one-loop effective potential in which the tree-level propagators are
replaced by temperature dependent effective propagators,
which were obtained by summing the dominant high temperature
contributions from  infinite-series of certain classes of self-energy
graphs in perturbation theory.
When one considers only the leading corrections to the effective
propagators  all results are in agreement with each other.
However, there have been various disagreements when the
subleading corrections to the effective propagators are included.
On the other hand, the subleading contributions
could be important in determining the nature of the phase transition.

We find that  in the improved one-loop calculations the difficulties
arise due to
the fact that the naive substitution of improved propagators in the
one-loop effective potential is an ad-hoc approximation. One needs a
self-consistent loop expansion of the effective potential in terms of
the full propagator.
Such a technique has been developed some time ago by Cornwall, Jackiw
and Tomboulis (CJT) in their effective action formalism for composite
operators${[\corn]}$. One considers a generalization $\Gamma (\phi,G)$
of the usual
effective action $\Gamma (\phi)$ which depends not only on $\phi(x)$
-a possible expectation value of the quantum field $\Phi(x)$- but also
on $G(x,y)$ -a possible expectation value of the time-ordered product
$T \Phi(x) \Phi(y)$. The physical solutions satisfy stationary
requirements
\eqn\stat{{\delta \Gamma(\phi,G) \over \delta \phi(x)} = 0 ~, }
\eqn\statg{{\delta \Gamma(\phi,G) \over \delta G(x,y)} = 0 . }

\vfill
\eject
The conventional effective action $\Gamma (\phi)$
is given by $\Gamma (\phi,G)$ at the solution $G_0(\phi)$ of \statg
\eqn\gusu{\Gamma(\phi) = \Gamma(\phi,G_{0}(\phi))  ~.}

In this formalism, it is possible to sum a large class of ordinary
perturbation-series diagrams that contribute to the effective action
$\Gamma (\phi)$, and the gap equation which determines the form of the
propagator  is obtained by a variational technique, as in \statg.

For translationally invariant solutions, we set
$\phi=$ constant, take $G(x,y)$ to be a function of only $(x-y)$, and
obtain the effective potential for the full propagator:
\eqn\veco{V(\phi,G) \equiv
\Gamma(\phi,G)^{trans. inv.} \bigg/  \int d^4x ~.}

The purpose of this paper is to understand the structure of the
leading corrections to the perturbative one-loop finite temperature
effective potential
for the $\lambda \Phi^4$ theory using CJT formalism in  imaginary
(Euclidean) time. We obtain a finite temperature effective potential,
which is exact up to the order that includes all contributions from
daisy and super-daisy graphs. Instead of dropping various  finite and
divergent terms, as has been done often in the recent literature, we
carry out renormalization and then perform a high temperature expansion.
We show explicitly that
the effective potential must be calculated up to
two-loop in order to generate all daisy and super-daisy
graphs which appear in perturbation theory$[\dine]$.
Moreover, we find subtle cancellations of leading corrections between the
improved one and two loop contributions. If it is indeed possible
to determine
the order of the electroweak phase transition
by calculating the improved high
temperature effective potential,
our result implies that the improved two-loop contribution could play a
crucial role. We plan to present our calculations
in gauge theories in a future publication.

\newsec{CJT Composite Operator Effective  Potential}
In this section we shall review briefly the CJT formalism following
Ref.[\corn]. To describe field theory at finite temperature T we shall
use Euclidean time $\tau$, which is restricted to the interval
$0 \leq \tau \leq \beta \equiv 1/T$. The Feynman rules are the same as
those at zero temperature, except that the momentum space integral over
the time component
$k_4$ is replaced by a  sum over discrete
frequencies $k_4 = 2 \pi n / \beta$ :
\eqn\disint{\int {d^4k \over (2 \pi)^4} \rightarrow {1 \over \beta}
\sum_n \int  {d^3k \over (2 \pi)^3}
{}~,}
where  $n$ is even (odd) for bosons (fermions).

We introduce a partition function $Z_\beta (J,K)$ in the presence of the
sources $J$ and $K$ defined by

\baselineskip 12pt plus 0.2pt minus 0.2pt
\eqnla\zcjt\zcjta{\eqalign{Z_\beta (J,K)  \equiv
\int D \Phi ~ exp - \biggl\{ &  I(\Phi) + \int d^4 x ~ \Phi(x) J(x) \cr
& + {1 \over 2} \int d^4 x ~ d^4 y ~ \Phi(x) K(x,y) \Phi(y) \biggr\} .}}
\baselineskip 24pt plus 0.2pt minus 0.2pt

\noindent
[We have set $c=\hbar=1$.]
The $\Phi$-integration is functional and $\Phi$ satisfies the periodic
boundary conditions $\Phi(\beta/2,{\bf x})=\Phi(-\beta/2,{\bf x})$.
$I(\Phi)$ is the classical Euclidean action, which may be written as

\baselineskip 12pt plus 0.2pt minus 0.2pt
\eqnlb\zcjtb{I(\Phi)= \int d^4 x ~ d^4 y ~ \Phi(x) D_0^{-1}(x-y)
\Phi(y) +  \int d^4 x ~ L_{int}(x)  , ~}

\noindent
where $D_0(x-y)$  is the free propagator

\eqnlcend\zcjtc{D_0^{-1}(x-y) =
- ( \square + m^2 ) \delta^4(x-y) ~,}
\baselineskip 24pt plus 0.2pt minus 0.2pt

\noindent
and the interaction Lagrangian $L_{int}$ is at least cubic in $\Phi$.

The effective action $\Gamma_\beta(\phi,G)$ is obtained by a double
Legendre
transformation of \space\space\space\space\space\space\space\space\space\space
$W_\beta(\phi,G) \equiv ln Z_\beta(\phi,G)$ which is
analogous to the free energy. We define
\eqn\fg{\eqalign{&{\delta W_\beta(J,K) \over \delta J(x)} = \phi(x)
{}~, \cr
&{\delta W_\beta(J,K) \over \delta K(x,y)} =
{1 \over 2} [G(x,y) + \phi(x) \phi(y)]
{}~, }}
and eliminate $J$ and $K$ in favor of $\phi$ and $G$
\eqn\gfg{\eqalign{\Gamma_\beta(\phi,G) & = W_\beta(J,K) -
\int d^4 x ~ \phi(x) J(x)  \cr
&~~ - {1 \over 2} \int d^4 x ~ d^4 y~ \phi(x) K(x,y) \phi(y)
- {1 \over 2} \int d^4 x ~ d^4 y~ G(x,y) K(y,x) ~. }}
It follows that
\eqn\invlega{{\delta \Gamma(\phi,G) \over \delta \phi(x)} =
J(x) - \int d^4 y ~K(x,y) \phi(y)  ~, }
\eqn\invlegb{{\delta \Gamma(\phi,G) \over \delta G(x,y)} =
- {1 \over 2} K(x,y) ~. }
Since the physical processes correspond to vanishing sources $J$ and
$K$, Eqns.\invlega\space and \invlegb\space provide the stationary requirement
of
\stat-\statg.
$\Gamma_\beta(\phi,G)$ obtained in this way is the generating functional
in $\phi$ for two-particle irreducible Green's functions expressed in
terms of the full propagator $G$.

In order to obtain a series expansion of $\Gamma_\beta(\phi,G)$ we
introduce the functional operator $D^{-1}(\phi;x,y)$:
\eqn\dfunct{D^{-1}(\phi;x,y) =
 {\delta^2 I(\phi) \over \delta \phi(x) \delta \phi(y)} ~.}
The required series obtained by CJT is then
\eqn\gexp{\Gamma_\beta(\phi,G)=I_{cl}(\phi)+{1 \over 2} TrLn D_0 G^{-1}
+{1 \over 2} Tr [ D^{-1} G - 1 ] + \Gamma_\beta^{(2)}(\phi,G)  ~,}
where the $\phi$-independent terms are chosen so that the overall
normalization is
consistent with the conventional effective action $\Gamma_\beta(\phi)$
according to \gusu. The quantity $\Gamma_\beta^{(2)}(\phi)$ is computed as
follows. In the classical action $I(\Phi)$, shift the field $\Phi$ by
$\phi$. Then $I(\Phi+\phi)$ contains terms cubic and higher in $\Phi$
that define $I_{int}(\phi;\Phi)$ where the vertices depend on $\phi$.
$\Gamma_\beta^{(2)}(\phi)$  is given by all the two-particle irreducible
(2PI) vacuum
graphs in the theory with vertices given by $I_{int}(\phi;\Phi)$ and
propagators set equal to $G(x,y)$.

{}From the stationary requirement \statg\space we obtain the gap equation for
$G$:
\eqn\gap{G^{-1}(x,y) =  D^{-1}(x,y)
- 2 {\delta \Gamma_\beta^{(2)}(\phi,G) \over \delta G(x,y)} ~.}

When one is interested in translationally invariant solutions, the
generalized effective potential $V_\beta(\phi,G)$ can be obtained using
\veco\space and \gexp.

\newsec{$\lambda \Phi^4$ Theory}

\centerline{{\bf A. Effective Potential $V_\beta(\phi,G)$}}

In this section we calculate the finite temperature effective potential
for the single  scalar field with $\lambda \Phi^4$ interaction. The
Euclidean Lagrangian density is given by
\eqn\lf{L = {1 \over 2} (\partial_{\mu} \Phi) (\partial^{\mu} \Phi)
+{1 \over 2} m^2 \Phi^2
+{\lambda \over 4!}  \Phi^4 ~.}
The propagator defined in \dfunct\space is
\eqn\dfif{D^{-1}(\phi;x,y) =
- ( \square + m^2 + {\lambda \over 2} \phi^2 )
\delta^4(x-y) ~,}
and the vertices of the shifted theory are given by
\eqn\sif{L_{int}(\phi;\Phi) =
{\lambda \over 6} \phi \Phi^3 + {\lambda \over 4!} \Phi^4 ~.}
In Fig.2 the diagrams contributing to $\Gamma^{(2)}_\beta(\phi,G)$ are shown
up to three loops: each line represents the propagator $G(x,y)$ and
there are two kind of vertices.

In order to determine how to truncate the loop expansion so that
all daisy and super-daisy graphs of
perturbation theory are included,  we  first observe that upon dropping
$\Gamma^{(2)}_\beta$
altogether, the gap equation \gap\space gives
\eqn\gapno{G^{-1}(x,y) =  D^{-1}(x,y) ~,}
and  $\Gamma(\phi,D)$  is simply the ordinary one-loop
effective potential.
Therefore, a  non-trivial $G$ can be obtained only if we retain some of
the graphs in $\Gamma^{(2)}_\beta$.
Next we observe that among the graphs in Fig. 2 only the two-loop
graph of $O(\lambda)$ will include contributions from daisy and
superdaisy graphs of ordinary perturbation theory. Therefore,
we shall truncate the series at $O(\lambda)$.
This is known as Hartree-Fock approximation.
The effective action is then
\eqn\gexpy{\Gamma_\beta(\phi,G)=I_{cl}(\phi)+{1 \over 2} TrLn D_0 G^{-1}
+{1 \over 2} Tr [ D^{-1} G - 1 ] +
{3 \over 4!} \lambda \int d^4x ~ G(x,x) G(x,x) ~.}
By stationarizing $\Gamma_\beta$ with respect to $G$ we obtain the gap
equation in the Hartree-Fock approximation:
\eqn\gapgfb{G^{-1}(x,y) = D^{-1}(x,y)
+ {\lambda \over 2}  G(x,x) \delta^4(x-y)  ~. }
It is straightforward to show by iteration that Eqn.\gapgfb\space
generates all daisy and super-daisy graphs
that contribute to the full propagator in ordinary
perturbation theory. [Eqns.\gexpy\space and \gapgfb\space have the same
structure as those in the leading large N approximation since in both
cases n-point functions are expressed in terms of one and two-point
functions. However, as discussed in Ref.[\corn],
in the
large N approximation those daisy and super-daisy diagrams which are
of $O(1/N)$ are dropped and therefore the coefficients of $\lambda$ in
\gexpy\space and \gapgfb\space
become smaller by a factor three.]

In order to obtain the effective potential $V_\beta(\phi,G)$ for
translation-invariant field configurations, we define the
Fourier-transformed propagators
\eqn\dftr{D(k)= \int {d^4k \over (2 \pi)^4} D(x-y) e^{i k (x-y)}=
{1 \over k^2 + m^2 + {\lambda \over 2} \phi^2}
{}~,}
\eqn\gftr{G(k)=\int {d^4k \over (2 \pi)^4} G(x-y) e^{i k (x-y)}=
{1 \over k^2 + M^2 }
{}~,}
where we have taken an {\it Ansatz} for $G(k)$ using an effective mass
$M^2$. Since the correction to the gap equation in this approximation is
given by $\lambda G(x,x)/2$, $M^2$ can be taken to be momentum
independent. The effective potential is then
\eqn\efpoa{\eqalign{V(\phi,M) =&  {1 \over 2} m^2 \phi^2
+{\lambda \over 4!}  \phi^4 +
\int {d^4k \over (2 \pi)^4}~
ln[k^2+M^2]  \cr
& - {1 \over 2 } [M^2-m^2- {\lambda \over 2} \phi^2] G(x,x)
+ {\lambda \over 8} G(x,x) G(x,x)     ~. }}
{}From \efpoa\space the stationary requirements are
\eqn\stataa{{\partial V_\beta(\phi,M) \over \partial \phi} =
\phi [m^2 + {\lambda \over 6} \phi^2 + {\lambda \over 2} G(x,x)] = 0 ~,}
\eqn\statbb{{\partial V_\beta(\phi,M) \over \partial M^2} =
-{1 \over 2} {\partial G(x,x) \over \partial M^2}
[M^2 - m^2 - {\lambda \over 2} \phi^2 - {\lambda \over 2} G(x,x)] = 0 ~.}

The conventional effective potential is obtained by evaluating
$V_\beta(\phi,M)$ at the solution ${M}(\phi)$ of \statbb.
It is composed of three terms: the classical ($V^0$), one-loop ($V^I$),
and two-loop ($V^{II}$) contributions.

\baselineskip 12pt plus 0.2pt minus 0.2pt
\eqnla\vsum\vsuma{V_\beta(\phi,{M}(\phi)) =V^{0}+
V^{I}+V^{II}
{}~,}
\eqnlb\vsumb{V^{0} = {1 \over 2} m^2 \phi^2
+{\lambda \over 4!}  \phi^4
{}~,}
\eqnlc\vsumc{V^{I} = \int {d^4k \over (2 \pi)^4}~
ln[k^2+{M}^2(\phi)]
{}~,}
\eqnld\vsumd{V^{II} = - {\lambda \over 8} G(x,x) G(x,x)
{}~,}
where
\eqnle\vsume{{M}^2(\phi) =
m^2 + {\lambda \over 2} \phi^2 + {\lambda \over 2} G(x,x)
{}~,}
and now $G(x,x)$ is given by
\eqnlfend\vsumf{G(x,x) = \int {d^4k \over (2 \pi)^4}~
{1 \over k^2 +  {M}^2(\phi) }
{}~.}
\baselineskip 24pt plus 0.2pt minus 0.2pt

\centerline{{\bf B.  Renormalizing the Effective Potential}}

The expression of $V_\beta(\phi,{M}(\phi))$ in \vsum\space contains
divergent integrals. Moreover, due to the fact that our
approximation is non-perturbatively self-consistent,
reflecting the non-linearity of
the full theory, ${M}(\phi)$, the argument of $V_\beta$, is not
well-defined due to the infinities in $G(x,x)$. Therefore, we shall
first obtain a well-defined, finite expression for ${M}(\phi)$ by a
renormalization.
[In the rest of this paper $M$ refers to the solution of \vsume.]
We define renormalized parameters $m_R$ and $\lambda_R$
as

\baselineskip 12pt plus 0.2pt minus 0.2pt
\eqnla\renml\renmla{\pm {m_R^2 \over \lambda_R} = {m^2 \over \lambda} +
{1 \over 2} I_1
{}~,}
\eqnlb\renmlb{{1 \over \lambda_R} = {1 \over \lambda} +
{1 \over 2} I_2(\mu)
{}~,}
\smallskip
\noindent
where $I_{1,2}$ are divergent integrals
\smallskip
\noindent
\eqnlc\renmlc{I_1 \equiv \int {d^3k \over (2 \pi)^3}~ {1 \over 2 |{\bf k}|} =
\lim_{\Lambda \rightarrow \infty} {\Lambda^2 \over 8 \pi^2}
{}~,}
\eqnldend\renmld{I_2 \equiv \int {d^3k \over (2 \pi)^3}~
[ {1 \over 2 |{\bf k}|} - {1 \over 2 \sqrt{|{\bf k}|^2 + \mu^2}} ] =
\lim_{\Lambda \rightarrow \infty} {1 \over 16 \pi^2}
\ln {\Lambda^2 \over \mu^2}
{}~.}
\baselineskip 24pt plus 0.2pt minus 0.2pt

\noindent
$\mu$ is the renormalization scale and $\Lambda$ is the ultraviolet
momentum cut-off. [The same renormalization prescription has been used
also in the large N approximation${[\momo]}$.]

When the sum on $n$ in $k_4$ is carried out as in Ref.[\doja], $G(x,x)$
becomes
\eqn\gxxb{\eqalign{G(x,x) &= \int {d^3k \over (2 \pi)^3}~
{1 \over 2 \omega_k } +
{1 \over \beta} \int {d^3k \over (2 \pi)^3}~
{1 \over \omega_k (exp[\beta \omega_k]-1)}   \cr
& \equiv G({M}(\phi)) + I_1 - {M}^2(\phi) I_2(\mu)  ~, }}
where $\omega_k \equiv [|{\bf k}|^2 + {M}^2(\phi)]^{1/2}$ and
$G({M}(\phi))$ is the finite part of $G(x,x)$ given by
\eqn\gom{G({M}(\phi)) \equiv  {{M}^2(\phi) \over 16 \pi^2}
\ln {{M}^2(\phi) \over \mu^2} +
{1 \over \beta} \int {d^3k \over (2 \pi)^3}~
{1 \over \omega_k (exp[\beta \omega_k]-1)}  ~.}
In the limit $T=0$, the first term in $G({M}(\phi))$ survives, but
the second term vanishes.

It is straightforward to see that ${M}^2(\phi)$ is finite and cut-off
independent in terms of $m_R$ and $\lambda_R$
\eqn\mgapfi{{M}^2(\phi) =
\pm m_R^2 + {\lambda_R \over 2} \phi^2
+ {\lambda_R \over 2} G({M}(\phi)) ~.}
[In this paper we shall choose the negative sign which
allows spontaneous symmetry breaking.] It is convenient for our later
discussions to define the tree-level effective mass $\tilde{m}^2(\phi)$
\eqn\mtrle{\tilde{m}^2(\phi) =
-m_R^2 + {\lambda_R \over 2} \phi^2
{}~.}

With this finite ${M}^2$, we are ready to discuss the
divergences in $V_\beta(\phi,{M})$. First, carrying out the sum
on $n$  in $V^I$, we obtain the familiar one-loop finite
temperature formula of Ref.[\doja], where the tree level effective
mass $\tilde{m}(\phi)$ is replaced by ${M}(\phi)$
\eqn\viom{\eqalign{V^I({M}) &= {1 \over 2} \int {d^3k \over (2 \pi)^3}~
 \omega_k +
{1 \over \beta} \int {d^3k \over (2 \pi)^3}~
\ln (1-exp[\beta \omega_k]) =  \cr
&=   {{M}^4 \over 64 \pi^2}
[\ln {{M}^2 \over \mu^2} - {1 \over 2}]
+ {1 \over \beta} \int {d^3k \over (2 \pi)^3}~
\ln (1-exp[\beta \omega_k])
- {{M}^4 \over 4}  I_2(\mu)
{}~. }}
At $T=0$ the first term  of $V^I$ survives and provides the zero-temperature
one-loop contribution, and the second term vanishes. The last term is the
divergence in $V^I$.

Divergences in the two-loop contribution $V^{II}$ come from $G(x,x)$ and
its square. Finiteness of $V_\beta(\phi,{M}(\phi))$ can be shown by
first combining $V^{0}$ and $V^{II}$ using the unrenormalized form of
the gap equation. When the combined expression is written in terms of
renormalized parameters, the remaining divergent integral is cancelled
by that of $V^{I}$ in \viom. This is another indication that the
two-loop contribution must be included for a finite self-consistent
approximation. The resulting finite expression for
$V_\beta(\phi,{M}(\phi))$ is

\baselineskip 12pt plus 0.2pt minus 0.2pt
\eqnla\vrensum\vrensuma{V_\beta(\phi,{M}(\phi)) = (V^{0}+V^{II})
+V^{I} ~,}
\eqnlb\vrensumb{V^{0}+V^{II} = {{M}^4 \over 2 \lambda_R}
- {1 \over 2} {M}^2 G({M})
- {\lambda \over 12}  \phi^4
{}~,}
\eqnlcend\vrensumc{V^{I} = {{M}^4 \over 64 \pi^2}
[\ln {{M}^2 \over \mu^2} - {1 \over 2}] +
{1 \over \beta} \int {d^3k \over (2 \pi)^3}~
\ln (1-exp[\beta \omega_k]) ~,}
\baselineskip 24pt plus 0.2pt minus 0.2pt

\noindent
[A constant term $m^4/(2 \lambda)$ has been adjusted to obtain
\vrensum\space from \vsum.] However, in order to compare the high
temperature effective potential with and without the two-loop
contribution in our later discussion, we still have to extract $V^{II}$
from \vrensumb. Observing that $V^0$ must be a function of
$\phi$ only, and
that in our approximation $V^{II}$ does not depend on $\phi$
explicitly [since
the graph of $O(\lambda)$ in Fig.2
does not involve any vertices which depend on
$\phi$] we obtain, by using the renormalized gap equation,
\eqn\vovii{V^{0}+V^{II} = \biggl[ {\lambda_R \over 8}
\biggl( \phi^2- 2 {m_R^2 \over \lambda_R} \biggr) -
{\lambda \over 12} \phi^4 \biggr] -
{\lambda_R \over 8} G({M}) G({M}) ~.}
Clearly the last term in \vovii\space is the two-loop contribution. The
quantity in the brackets is the classical contribution after
renormalization is carried out. It is cut-off dependent
because of
the term $- \lambda \phi^4 /12$, which did not get renormalized due
to the structure of the gap equation.
But
the renormalization prescription \renml\space tells us that if
$\lambda_R$ is held fixed as $\Lambda \rightarrow \infty$, $\lambda$
approaches $0_-$. [A necessary condition in a renormalized $\lambda
\phi^4$ theory is $\lambda < 0$.] As shown in the large N
studies${[\momo]}$, such theory is intrinsically unstable. On the other
hand, holding $\lambda > 0$ implies $\lambda_R \rightarrow 0$ as
$\Lambda \rightarrow \infty$. For $\lambda > 0$, a sensible
theory can be obtained for a fixed small $\lambda_R > 0$ as an
effective low energy theory, if $\Lambda$ is kept fixed at a large but
finite value.
Such theory requires
\eqn\condapi{{\lambda_R \over 32 \pi^2} \ln {\Lambda^2 \over \mu^2} < 1
{}~,}
in order to have $\lambda >0$, and all momenta, temperature and any
other physical mass scale must be much smaller than $\Lambda$.

As shown in Fig.3, the zero-temperature phase structure of the effective
theory with finite $\Lambda$ is similar to that of perturbation theory:
there exists a minimum at a non-zero value of $\phi$${[\pisa]}$.
We shall consider such an effective theory and take our temperature to
be $T<<\Lambda$.

\centerline{{\bf C. High Temperature Effective Potential}}

Our main interest is in the form of the high temperature effective
potential. We first find the high temperature gap equation by expanding
the integral expression of $G({M})$, i.e. the second term in \gom\space
at high temperature. Since
the basic mass scale in the problem is ${M}$, we consider an
expansion in ${M}^2/T^2 << 1$
\eqn\gomt{G({M}) = T^2
\biggl[{1 \over 12 } - { 1 \over 4 \pi} {{M}  \over T } +
O \biggl( {{M}^2 \over T^2} \ln T \biggr) \biggr] ~. }
[We have chosen our renormalization scale $\mu$ to be $m_R$.]
Then the high temperature gap equation
takes the form

\baselineskip 12pt plus 0.2pt minus 0.2pt
\eqn\mgapfit{{M}^2= \tilde{m}^2(\phi)
+ {\lambda_R \over 24} T^2 -
{\lambda_R \over 8 \pi} {M} T
+O \biggl( \lambda_R {M}^2 \ln T \biggr)
{}~.}
\baselineskip 24pt plus 0.2pt minus 0.2pt

\noindent
{}From the solution of this equation one finds that for a small coupling
$\lambda_R << 1$, the condition ${M}^2/T^2 << 1$ is consistent
with
\eqn\condbpi{\tilde{m}^2(\phi)/T^2 << 1 ~,}
which is exactly the required condition for high temperature expansion
of the perturbative calculation in Ref.[\doja].

Now we return to Eqn.\vrensum\space and \vovii. The high temperature
expansion of the
effective potential can be obtained using the high temperature expansion
of $G({M})$ and also the high temperature expansion of the
perturbative one-loop effective potential of Ref.[\doja] by replacing
the tree-level effective mass $\tilde{m}(\phi)$ by ${M}$

\baselineskip 12pt plus 0.2pt minus 0.2pt
\eqnla\rew\rewa{V_\beta(\phi,{M}(\phi))= V^{0}+V^{I}
+V^{II}
{}~, ~~~~~~~~~~~~~~~~~~~~~~}
\eqnlb\rewb{V^0 = {\lambda_r \over 8}
\biggl[\phi^2- 2 {m_R^2 \over \lambda_R}
\biggr]^2 - {\lambda \over 12}  \phi^4
{}~, ~~~~~~~~~~~~~~~~~~~~~~ }
\eqnlc\rewc{
V^{I} = - {\pi^2 \over 90} T^4 +
{{M}^2 T^2 \over 24} -
{{M}^3 T \over 12 \pi}
+ O({M}^4 \ln {{M}^2 \over T^2})
{}~,}
\eqnldend\rewd{V^{II}= - {\lambda_R \over 8}  \biggl[
 {T^4 \over 144}  -
{{M} T^3 \over 24 \pi}
+ {{M}^2 T^2 \over 16 \pi^2}
+ O({M}^4 \ln {{M}^2 \over T^2}) \biggr] ~.}
\baselineskip 24pt plus 0.2pt minus 0.2pt

\newsec{Discussions and Conclusions}

In order to understand the structure of the effective potential in
our approximation, we shall first consider the non-linear
aspects of the gap equation \gapgfb, which in the high temperature limit
can be expressed as
\eqn\congap{{M}^2(\phi) = \tilde{m}^2(\phi)
+ {\lambda_R \over 24} T^2
- {\lambda_R \over 8 \pi}  T {M}(\phi)
{}~.}
Consequently, ${M}(\phi)$ can be expanded
for small $\lambda_R$ as

\baselineskip 18pt plus 0.2pt minus 0.2pt
\eqnla\maexp\maexpa{{M}(\phi) = M_L(\phi)
\biggl\{ 1 - {\lambda_R T \over 16 \pi M_L(\phi) } +
O\biggl[ \biggl({ \lambda_R T \over 16 \pi M_L(\phi)} \biggr)^2 \biggr]
\biggr\}
{}~,}

\vskip 0.5 truecm
\noindent
where

\eqnlbend\maexpb{M_L(\phi) \equiv \sqrt{\tilde{m}^2(\phi)
+ {\lambda_R \over 24} T^2}
{}~}
\baselineskip 24pt plus 0.2pt minus 0.2pt

\noindent
solves the linearized high temperature gap equation, i.e. \congap\space
without the last term.

When the one-loop contribution $V^I$ in \rewc\space is rewritten using
the gap equation, we obtain
\eqn\vonma{V^{I}(\phi) = - {\pi^2 \over 90} T^4 +
{M_L^2(\phi) T^2 \over 24} -
{\lambda_R  \over 192 \pi } {M}(\phi) T^3 -
{{M}^3(\phi) T \over 12 \pi}
+ O({{M}^4 \over 64 \pi^2 } \ln T) ~.}

We observe that the term linear in ${M}$, namely the third term on
the right-hand side of Eqn.\vonma, arises from the non-linearity of the
gap equation, i.e. from the last term in \congap. If we were to use the
linearized gap equation, without this term,
the first non-trivial correction to the
perturbative one-loop effective potential would be given by the term cubic in
${M}(\phi)$. However,
at high temperatures the leading non-linear correction is of the same
order as the term cubic in
${M}(\phi)$ in \vonma; in fact from \congap\space we have
\eqn\blublu{\left({\lambda_R  \over 192 \pi } {M}(\phi) T^3 \right) \bigg/
\left({{M}^3(\phi) T \over 12 \pi}\right) \simeq {288 \over 192} \sim O(1)
{}~ }
for $T >> \phi$, and such a term could alter the nature of
the phase transition. When we include the two-loop contribution given in
\rewd\space the ${M} T^3 $ term disappears and the high
temperature effective potential in our approximation is
(neglecting $\phi$-independent contributions)
\eqn\gacpihf{V(\phi) = V^{0}(\phi) +
\biggl({ T^2 \over 24} M_L^2(\phi) -
{T \over 12 \pi } M_L^3(\phi) \biggr) \biggl(1 + O(\lambda_R) \biggr)
+ O({{M}^4 \over 64 \pi^2 } \ln T) ~.}
In Fig.4,
the  effective potential of
Eqn.\gacpihf\space is shown as a function of $\phi$ for five different
temperatures close to
the critical temperature. It is evident that there is a temperature
such that there are two degenerate minima.

The above analysis of our consistent approximation shows that
improving the perturbative one-loop effective potential using the
non-linear gap equation clearly leads to an erroneous
result${[\ref\notarn{P. Arnold in Ref.[\arn] pointed this out, using the
language of ordinary peturbation theory.}]}$.
Therefore, one must use a self-consistent method which relates the
effective potential and the gap equation.
On the other hand, we also
see that, due to the cancellation of the leading non-linear effect,  one
can in fact obtain a consistently improved effective
potential by improving the perturbative one-loop effective potential
using an effective mass $M_L(\phi)$, which is the solution of the
linearized gap equation. Such
improved perturbation theory, using the effective mass squared shifted by
a $\phi$-independent amount proportional to $T^2$, was first suggested
by Weinberg$[\wei]$ and later further studied by others${[\ref\fen{P.
Fendley, Phys. Lett. {\bf B 196}, 175 (1987).}]}$.
If the cancellation of the leading non-linear effects in
our approximation is a general feature, occurring even in gauge
theories, the one-loop improved effective potential in the standard
model first calculated by Carrington$[\car]$ would be a consistent
approximation. We plan to clarify this in a future publication.

\vskip1.5cm

\centerline{{\bf ACKNOWLEDGEMENTS}}

One of us (G. A.-C.) acknowledges profitable
discussions with S.D.H. Hsu and T.S. Evans.
G. A.-C. acknowledges support from the ``Fondazioni
Angelo Della
Riccia'', Firenze, Italy.

\vfill
\eject
\vfill
\vfill\eject\immediate\closeout\rfile
\centerline{{\bf References}}\bigskip{
\catcode`\@=11\escapechar=` %
\input refs.tmp\vfill\eject}
\vfill
\eject
\vfill
\figures
\fig1 Examples of daisy (a) and superdaisy (b) graphs.
\vskip0.8cm
\fig2 The two- and three- loop  graphs contributing to
$\Gamma_\beta^{(2)}(\phi,G)$. They are two-particle irreducible and their lines
represent the full propagator G.
\vskip0.8cm
\fig3 $V_\beta(\phi,{M}(\phi)) / m_R^4$ as a function of $\phi / m_R$ at
$T=0$. $V_\beta$ illustrated in figure corresponds
to $\lambda_R = 0.05$ and $\ln (\Lambda^2/ m_R^2) = 16 \pi^2$.
$V_\beta$ becomes imaginary for small $\phi / m_R$.
\vskip0.8cm
\fig4 The high temperature effective potential in our
approximation (Eqn.\gacpihf) for five different temperatures close to
the critical temperature. [Here we have chosen
 $\lambda_R = 0.05$ and $\ln (\Lambda^2/ m_R^2) = 16 \pi^2$.] In order
to  compare the shapes between different temperatures, we shifted the
$V_\beta$'s by
$\phi$-independent amounts so that $V_\beta=0$ for $\phi / m_R = 0$.
{}From the figure it is evident that at a certain temperature there are
two degenerate minima.

\end